\documentclass[12pt]{article}
\usepackage{graphicx}
\usepackage{epstopdf}
\usepackage{caption}
\usepackage{subcaption}
\usepackage{subfig}

\usepackage{epsfig,amsmath,amssymb,verbatim,mathrsfs}

\def\beq{\begin{eqnarray}}
\def\eeq{\end{eqnarray}}
\def\bea{\begin{eqnarray}}
\def\eea{\end{eqnarray}}
\newcommand{\s}{\text{\tiny S}}
\newcommand{\h}{\text{\tiny H}}

\newcommand{\dm}{\text{\tiny DM}}

\renewcommand{\thefootnote}{\roman{footnote}}

\begin{document}

\setlength{\baselineskip}{0.2in}


\begin{titlepage}
\noindent
\flushright{April 2015}
\vspace{0.2cm}

\begin{center}
  \begin{Large}
    \begin{bf}
The Scale-Invariant Scotogenic Model\\

     \end{bf}
  \end{Large}
\end{center}

\vspace{0.2cm}

\begin{center}

\begin{bf}

{Amine~Ahriche,$^{1,2,}$\footnote{aahriche@ictp.it}
Kristian~L.~McDonald$^{3,}$\footnote{kristian.mcdonald@sydney.edu.au} and
Salah~Nasri$^{4,5,}$\footnote{snasri@uaeu.ac.ae}}\\
\end{bf}
\vspace{0.5cm}
 \begin{it}
$^1$ Department of Physics, University of Jijel, PB 98 Ouled Aissa, DZ-18000 Jijel, Algeria\\
\vspace{0.1cm}
$^2$ The Abdus Salam International Centre for Theoretical Physics, Strada Costiera 11, I-34014, Trieste, Italy\\
\vspace{0.1cm}
$^3$ ARC Centre of Excellence for Particle Physics at the Terascale,\\
School of Physics, The University of Sydney, NSW 2006, Australia\\
\vspace{0.1cm}
$^4$ Physics Department, UAE University, POB 17551, Al Ain, United Arab Emirates \vspace{0.3cm}
\end{it}
\vspace{0.5cm}

\end{center}


\begin{abstract}

We investigate a minimal scale-invariant implementation of the scotogenic model and show that  viable electroweak symmetry breaking  can occur while simultaneously generating one-loop neutrino masses and the dark matter relic abundance. The model predicts the existence of a singlet scalar (dilaton) that plays the dual roles of triggering electroweak symmetry breaking and sourcing lepton number violation. Important constraints are studied, including those from lepton flavor violating effects and dark matter direct-detection experiments. The latter turn out to be somewhat severe, already excluding large regions of parameter space. None the less, viable regions of parameter space are found, corresponding to dark matter masses  below (roughly) 10~GeV and above 200 GeV.

\end{abstract}

\vspace{1cm}

\end{titlepage}
\renewcommand{\thefootnote}{\arabic{footnote}}
\setcounter{footnote}{0}
\setcounter{page}{1}


\vfill\eject


\section{Introduction\label{sec:introduction}}

The discovery of the Higgs boson  provides an explanation for the origin of mass in the charged fermion and gauge sectors of the Standard Model (SM). However, despite this great success, a number of problems remain. In particular, our understanding of the origin of neutrino mass is incomplete, and we do not know the constituent properties of the dark matter (DM) that appears necessary on galactic scales. In addition to these puzzles, the origin of the $\mathcal{O}(100)$~GeV mass-parameter that determines the weak scale in the SM also remains a mystery. Thus, with regard to the mechanisms of mass in the universe, there remains much to be discovered.

The scotogenic model is a simple framework that aims to address some of these short-comings~\cite{Ma:2006km}. It offers an explanation for the origin of neutrino mass and the nature of DM by proposing a common or unified solution to these puzzles. In this approach, neutrinos acquire mass as a radiative effect, at the one-loop level, due to interactions with a $Z_2$-odd sector that includes DM candidates. The resulting theory gives a simple model for neutrino mass and DM, and has been well-studied in the literature~\cite{Schmidt:2012yg}.

Motivated by  the simplicity of the scotogenic model, and our inadequate understanding of the origin of the weak scale, in this work  we investigate a minimal scale-invariant (SI) implementation of the scotogenic model (hereafter, the SI scotogenic model). Our goal is to maintain the appealing features of the scotogenic model, namely the explanation for both neutrino mass and DM, while incorporating a dynamical model for the origin of the weak scale. In such a model, the dimensionful parameters, including the Higgs mass, are born as a dynamical effect via  radiative symmetry breaking~\cite{Coleman:1973jx}.  Due to their common origin,  both the Higgs mass and the exotic masses should appear at a similar scale, of $\mathcal{O}(\mathrm{TeV})$, enhancing the prospects for testing the model. The resulting theory provides a common framework for the aforementioned problems relating to mass - namely the origin of neutrino mass, the origin of the weak scale, and the nature of DM. 

We investigate the SI scotogenic model in detail, demonstrating that viable electroweak symmetry breaking can be achieved, while simultaneously generating neutrino masses and the DM relic abundance. The model predicts a singlet scalar (dilaton) that plays two important roles - it triggers electroweak symmetry breaking and sources the lepton number violation that allows radiative  neutrino mass. Important constraints are studied, including those from lepton flavor violating effects, DM direct-detection experiments, and the Higgs sector, such as  the invisible Higgs decay width and Higgs-dilaton mixing. Direct-detection constraints turn out to be rather severe and we find that large regions of parameter space are already excluded. None the less, viable parameter space  is found with a DM mass below (roughly) 10~GeV or above 200~GeV. The model can be experimentally probed in a number of ways, including: $\mu\rightarrow e+\gamma$ searches, future direct-detection experiments,  precision studies of the Higgs decays $h\rightarrow \gamma\gamma$ and $h\rightarrow \gamma Z$, and collider searches for an inert doublet.

Before proceeding we note that  a number of earlier  papers have studied relationships between neutrino mass and DM; see e.g.~Refs.~\cite{Krauss:2002px,Aoki:2013gzs,Ng:2014pqa,Culjak:2015qja}, and also Ref.~\cite{Ahriche:2015wha}, in which DM stability follows from an accidental symmetry. Earlier works investigating SI extensions of the SM appear in Ref.~\cite{Hempfling:1996ht} and, in particular, studies of SI models for neutrino mass can be found  in Refs.~\cite{Foot:2007ay,Ahriche:2015loa}. 

The structure of this paper is as follows. In Section~\ref{sec:scotogenic_SI} we introduce the model and detail the symmetry breaking sector. We turn our attention to the origin of neutrino mass in Section~\ref{sec:neutrino_mass} and discuss various constraints in Sections~\ref{sec:higgsdecay} and~\ref{sec:constraints}. Dark matter is discussed in  Section~\ref{sec:dark_matter} and our main analysis and results appear in Section~\ref{sec:results}. Conclusions are drawn in Section~\ref{sec:conc}.


\section{The Scale-Invariant Scotogenic Model\label{sec:scotogenic_SI}}

The minimal SI implementation of the scotogenic model is obtained by extending the SM to include three generations of gauge-singlet fermions, $N_{iR}\sim(1,1,0)$, where $i=1,\,2,\,3,$ labels generations, a second SM-like scalar doublet, $S\sim(1,2,1)$, and a singlet  scalar $\phi\sim(1,1,0)$. A $Z_2$ symmetry with action $\{N_R,\, S\}\rightarrow - \ \{N_R,\, S\}$ is imposed on the model.\footnote{This model was also mentioned in Ref.~\cite{Lee:2012jn}.}  The scalar $\phi$, as well as the SM fields,  transform trivially under this symmetry. The lightest $Z_2$-odd particle is stable and may be a DM candidate; this should be taken as either the lightest singlet fermion $N_{1}$ or a neutral component of the the doublet $S$, as discussed below. The scalar $\phi$ plays the dual roles of sourcing lepton number violation, to allow neutrino mass, and triggering electroweak symmetry breaking.

With this field content,  the most-general Lagrangian consistent with both the SI and $Z_2$ symmetries contains the terms 
\bea
\mathcal{L}&\supset& i\bar{N_R}\gamma^\mu \partial_\mu N_R +\frac{1}{2}(\partial^\mu \phi)^2+|D^\mu S|^2 - \frac{y_i}{2}\;\phi\,\overline{N_{iR}^c}\,N_{iR} -g_{i\alpha} \overline{N_{iR}}\,L_\alpha S-V(\phi,S,H),\label{eq:lagrange}
\eea
where $L_\alpha \sim(1,2,-1)$ denotes the SM lepton doublets, with generations labeled by Greek letters, $\alpha,\,\beta=e,\,\mu,\,\tau$. We denote the SM scalar doublet  as $H\sim(1,2,1)$  and  $V(\phi,S,H)$ is the most-general scalar potential consistent with the symmetries. The SI symmetry precludes any dimensionful parameters in the model, including bare Majorana mass terms for the fermions $N_i$. 

\subsection{Symmetry Breaking}

In the absence of dimensionful parameters,  the scalar potential contains only quartic interactions:
\bea
V(\phi,S,H)&=& \lambda_\h|H|^4+\frac{\lambda_\phi}{4} \phi^4 +\frac{\lambda_\s}{2}|S|^4 +\frac{\lambda_{\phi\h}}{2} \phi^2 |H|^2+\frac{\lambda_{\phi\s}}{2} \phi^2 |S|^2+\lambda_{3} |H|^2 |S|^2\nonumber\\
& &  +\lambda_{4}\,|H^\dagger S|^2 +\frac{\lambda_{5}}{2}(S^\dagger H)^2+\mathrm{H.c.} \label{eq:SIma_potential}
\eea
where $\lambda_5$ can be taken real without loss of generality. The desired VEV pattern has $\langle S\rangle=0$, to preserve the $Z_2$ symmetry, with $\langle H\rangle\ne0$ and $\langle\phi\rangle\ne0$, to break both the SI and electroweak symmetries. In addition to the doublet scalar $S$, we shall see that the spectrum contains an SM-like scalar $h_1$ and a dilaton $h_2$. 

Radiative corrections play an important role in triggering the desired symmetry breaking pattern. A full analysis of the potential requires the inclusion of leading-order loop corrections; however, in general, the full one-loop corrected potential is not analytically tractable. None the less, as discussed in Ref.~\cite{Ahriche:2015loa} (and guided by Ref.~\cite{Gildener:1976ih}), simple analytic expressions can be obtained by noting the following. Loop corrections involving SM fields are dominated by top-quark loops, due to the large Yukawa coupling. To allow viable electroweak symmetry breaking and give a positively-valued dilaton mass, these corrections must be dominated by loop corrections from a beyond-SM scalar, namely $S$. Thus, loop corrections from $S$ and $t$ are expected to dominate and,  to reasonable approximation, one can neglect loop corrections involving the light scalars (namely  the SM-like Higgs and the dilaton). More precisely, this gives an approximation to the potential up to corrections of $\mathcal{O}(M_{h_{1}}^4/M_S^4)$~\cite{Ahriche:2015loa}, which is reasonable provided one restricts attention to $M_S\gtrsim 200$~GeV.

Adopting this approximation, and writing the SM scalar in unitary gauge as $H=(0,\, h/\sqrt{2})$, the one-loop corrected potential for $h$ and $\phi$ is 
\begin{eqnarray}
V_{1-l}\left( h,\phi \right) &=&\frac{\lambda _{\h}}{4}h^{4}+\frac{\lambda _{\phi \h}}{4}\phi ^{2}h^{2}+\frac{\lambda
_{\phi }}{4}\phi ^{4}+\sum_{i=all~fields}n_{i}\,G\left( M_{i}^{2}\left( h,\phi
\right) \right), \label{V}
\end{eqnarray}
where $n_i$ is a multiplicity factor, $\Lambda$ is the renormalization scale, and the sum is over all fields barring the light scalars ($h$ and $\phi$) and the light SM fermions (all but the top-quark). The function $G$ is given by
\bea
G\left( X \right) &=&\frac{X ^{2}}{64\pi ^{2}}\left[ \log \frac{X }{\Lambda ^{2}}-\frac{3}{2}\right].
\eea
In the absence of bare dimensionful parameters, the field-dependent masses can be written as
\begin{equation}
M_{i}^{2}\left( h,\phi \right) =\frac{\alpha _{i}}{2}h^{2}+\frac{\beta _{i}}{2}\phi ^{2},
\end{equation}%
where $\alpha _{i}$ and $\beta _{i}$ are  constants.

Symmetry breaking is triggered via dimensional transmutation, introducing a dimensionful parameter into the theory in exchange for one of the dimensionless couplings (which is now fixed in terms of the other parameters). Analyzing the potential reveals a minimum with both $\langle \phi \rangle \equiv x\neq 0$ and $\langle h\rangle \equiv v\neq 0$ for $\lambda _{\phi \h}<0$. If one considers the tree-level potential, the desired VEV pattern is triggered at the scale $\Lambda$ where the running couplings obey $2\sqrt{\lambda_\h(\Lambda)\lambda_\phi(\Lambda)} +\lambda_{\phi\h}(\Lambda)=0$. Including loop corrections, subject to our approximation, modifies this relation to 
\begin{equation}
2\left\{ \lambda _{ \h}{\lambda _{\phi }}+\frac{\lambda _{\h}\ }{x^{2}}\sum_{i}n_{i}\left\{ \beta _{i}-\alpha _{i}\frac{v^{2}}{%
x^{2}}\right\} G^{\prime }\left( M_{i}^{2}\right) \right\}
^{1/2}+\lambda _{\phi \h}+\frac{2}{x^{2}}\sum_{i}n_{i}\alpha _{i}G^{\prime }\left(
M_{i}^{2}\right) =0, \label{1loopcoupling_condition}
\end{equation}%
with $G^{\prime }\left( \eta \right) =\partial G\left( \eta \right)
/\partial \eta $. The further condition%
\begin{equation}
-\frac{\lambda _{\phi \h}}{2\lambda _{\h}}=\frac{%
v^{2}}{x^{2}}+\sum_{i}\frac{n_{i}\alpha _{i}}{\lambda _{\text{{\tiny
H}}}\ x^{2}}G^{\prime }\left( M_{i}^{2}\right), \label{eq:1loopvev_cond}
\end{equation}%
is also satisfied. Absent fine-tuning, we observe that with $\lambda _{\h,\phi\h}=\mathcal{O}(1)$ one obtains $v\sim x$ and the exotic scale is
 expected near the TeV scale.  Eqs.~%
\eqref{1loopcoupling_condition} and \eqref{eq:1loopvev_cond} ensure that the tadpoles
vanish.

One-loop vacuum stability  requires that the couplings obey:
\begin{equation}
\lambda _{\text{{\tiny H}}}^{1-l},\lambda _{\phi }^{1-l},\lambda
_{\phi \text{{\tiny H}}}^{1-l}+2\sqrt{\lambda _{\text{{\tiny
H}}}^{1-l}\lambda _{\phi }^{1-l}}>0, \label{1loopstability_condition}
\end{equation}%
where the one-loop couplings are defined as
\begin{equation}
\lambda _{\h}^{1-l}=\frac{1}{6}\frac{\partial
^{4}V_{1-l}}{\partial h^{4}},\quad \lambda _{\phi }^{1-l}=\frac{1}{6}\frac{\partial ^{4}V_{1-l}}{\partial \phi
^{4}},\quad \quad \lambda _{\phi \h
}^{1-l}=\frac{\partial ^{4}V_{1-l}}{\partial h^{2}\partial \phi ^{2}}.
\end{equation}%
Eq.~\eqref{1loopstability_condition}  guarantees
that the masses for the neutral scalars $h$ and $\phi$ are strictly
positive, forcing one of the beyond-SM scalars in the doublet $S$ to be the
heaviest particle in the spectrum, to overcome top-quark contributions to the dilaton mass. Demanding $\lambda _{\phi \h}^{1-l}<0$
also ensures that the vacuum with $v\neq 0$ and $x\neq 0$ is preferred over the vacuum with a single nonzero VEV.

\subsection{The Scalar Spectrum}

Writing the inert-doublet as $S=(S^{+},(S^{0}+iA)/\sqrt{2})^{T}$, the
components have masses
\bea
M_{S^{+}}^{2}&=&\frac{\lambda _{\phi \text{{\tiny S}}}}{2}\,x^{2}+\frac{%
\lambda _{3}}{2}\,v^{2},\nonumber\\
M_{S^{0},A}^{2}&=&\frac{\lambda _{\phi \text{{\tiny S}%
}}}{2}\,x^{2}+(\lambda _{3}+\lambda _{4}\pm \lambda _{5})\frac{v^{2}}{2}\ =\  M_{S^{+}}^{2}+(\lambda _{4}\pm \lambda _{5})\frac{v^{2}}{2}.
\eea
The $\lambda _{5}$-term splits the neutral scalar masses $%
M_{S^{0}}$ and $M_{A}$, with the splitting becoming negligible in the limit $%
\lambda _{5}\ll 1$.\footnote{Note that the limit $\lambda_5\ll1$ is technically natural due to the restoration of lepton number symmetry in the limit $\lambda_5\rightarrow0$.} After symmetry breaking, the scalars $h$ and $\phi $ mix
to give two mass eigenstates, which we denote by $h_{1,2}$,
\begin{equation}
h_{1}\,=h\,\cos \theta _{h}\,-\phi \sin \theta _{h}\,\,,\quad
\,h_{2}=h\,\sin \theta _{h}\,+\phi \cos \theta _{h}\,\,.
\label{eq:1loop_scalar_eigenstates}
\end{equation}%
Due to the $Z_{2}$ symmetry, the neutral components of $S$ do not mix with
these fields. At tree-level the mixing angle is determined by the VEVS,
\begin{equation}
c_{h}\ \equiv \ \cos \theta _{h}\ =\ \frac{x}{\sqrt{x^{2}+v^{2}}},\quad
s_{h}\ \equiv \ \sin \theta _{h}\ =\ \frac{v}{\sqrt{x^{2}+v^{2}}},\,
\end{equation}%
and the SM-like scalar mass is given by
\begin{equation}
M_{h_{1}}^{2}=(2\lambda _{\text{{\tiny H}}}\ -\lambda _{\phi \text{{\tiny H}}%
})v^{2}\ \simeq \ 125~\mathrm{GeV}.
\end{equation}%
The scalar $h_{2}$ is the pseudo-Goldstone boson associated with the broken
SI symmetry, and is massless at tree-level, though radiative corrections
induce $M_{h_{2}}\neq 0$. A useful approximation for $M_{h_{2}}$ is~\cite%
{Gildener:1976ih}
\begin{equation}
M_{h_{2}}^{2}\simeq \frac{1}{8\pi ^{2}(x^{2}+v^{2})}\left\{
M_{h_{1}}^{4}+6M_{W}^{4}+3M_{Z}^{4}-12M_{t}^{4}+2M_{S^{+}}^{4}+M_{A}^{4}+M_{S^{0}}^{4}-2\sum_{i=1}^{3}M_{%
{i}}^{4}\right\}. \label{eq:pgb_mass}
\end{equation}%
Here the singlet fermion masses are given by $M_{i}=y_{i}\,x$, and are
ordered as $M_{1}<M_{2}<M_{3}$. Eq.~\eqref{eq:pgb_mass} shows that viable symmetry breaking requires one of the scalars $S^+$, $S^0$ or $A$ to be the
heaviest particle in the spectrum, to overcome negative loop
contributions to $M_{h_{2}}$ from the top quark and the fermions $N_{i}$.

Tree-level expressions for $M_{h_{1}}$ and $\theta
_{h}$ are presented above for convenience, however, in our numerical analysis (detailed below), we
use the mass eigenvalues $M_{h_{1,2}}$ and the mixing angle $\theta _{h}$
obtained by diagonalizing the one-loop corrected potential. We note that the SI
symmetry imposes non-trivial constraints on the model, with $\lambda _{\phi
} $ and $\lambda _{\phi \text{{\tiny H}}}$ fixed by Eqs.~%
\eqref{1loopcoupling_condition} and \eqref{eq:1loopvev_cond}, and the Higgs
mass $M_{h_{1}}\simeq 125$~\textrm{GeV} further fixes $\lambda _{\text{{\tiny H}}}$.

\section{Neutrino Mass\label{sec:neutrino_mass}}

\begin{figure}[ttt]
\begin{center}
        \includegraphics[width = 0.50\textwidth]{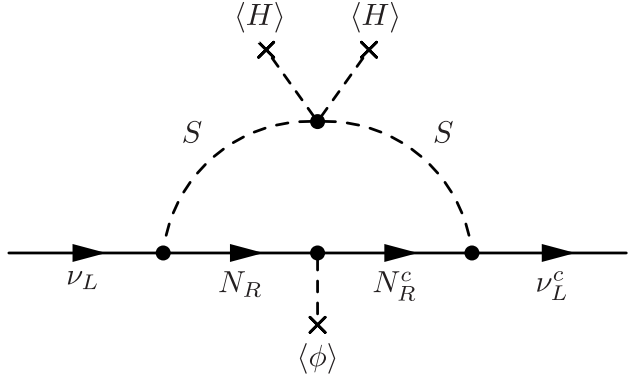}
\end{center}
\caption{One-loop diagram for neutrino mass in the scale-invariant scotogenic model.}\label{fig:SI_one_loop}
\end{figure}

The combined terms in Eqs.~\eqref{eq:lagrange} and~\eqref{eq:SIma_potential}
explicitly break lepton number symmetry, giving rise to radiative neutrino mass at the one-loop level, as shown in Figure~\ref{fig:SI_one_loop}. Observe that $\phi$ plays a key role in allowing the neutrino mass diagram, without which neutrinos would remain massless.\footnote{The Feynman diagram in Figure~\ref{fig:SI_one_loop} is an example of the SI type T3 one-loop topology. Related variants are possible~\cite{in_prep}.}
Calculating the mass diagram gives
\begin{equation}
(\mathcal{M}_{\nu })_{\alpha \beta }=\sum_{i}\frac{g_{i\alpha }g_{i\beta
}M_{i}}{16\pi ^{2}}\left\{ \frac{M_{S^{0}}^{2}}{M_{S^{0}}^{2}-M_{i}^{2}}\ln
\frac{M_{S^{0}}^{2}}{M_{i}^{2}}-\frac{M_{A}^{2}}{M_{A}^{2}-M_{i}^{2}}\ln
\frac{M_{A}^{2}}{M_{i}^{2}}\right\} .
\end{equation}%
In the limit that $M_{S^{0}}^{2}\approx M_{A}^{2}\equiv M_{0}^{2}$, this
simplifies to
\begin{equation}
(\mathcal{M}_{\nu })_{\alpha \beta }\simeq \sum_{i}\frac{g_{i\alpha
}g_{i\beta }\lambda _{5}v^{2}}{16\pi ^{2}}\frac{M_{i}}{M_{0}^{2}-M_{i}^{2}}%
\left\{ 1-\frac{M_{i}^{2}}{M_{0}^{2}-M_{i}^{2}}\ln \frac{M_{0}^{2}}{M_{i}^{2}%
}\right\} .
\end{equation}%
Note that the $Z_{2}$ symmetry prevents mixing between SM neutrinos and the
exotics $N_{i}$.

One can relate the entries in the neutrino mass matrix to the elements of the Pontecorvo-Maki-Nakawaga-Sakata (PMNS) mixing matrix~\cite{Pontecorvo:1967fh} elements. We parameterize the latter as
\bea
U_{\nu }=\left(
\begin{array}{ccc}
c_{12} c_{13} & c_{13} s_{12} & s_{13}e^{-i\delta_{d}} \\
-c_{23} s_{12}-c_{12} s_{13}s_{23}e^{i\delta_{d}}&c_{12} c_{23}-s_{12} s_{13}s_{23}e^{i\delta_{d}} &c_{13}s_{23} \\
s_{12} s_{23}-c_{12} c_{23}s_{13}e^{i\delta_{d}}&-c_{12} s_{23}-c_{23} s_{12}s_{13}e^{i\delta_{d}} &c_{13}c_{23}
\end{array}
\right) \times U_{m},
\eea
with $\delta _{d}$ being the Dirac phase and $U_{m}=\mathrm{diag}(1,\,e^{i\theta _{\alpha }/2},\,e^{i\theta _{\beta
}/2})$ giving the dependence on the Majorana phases $\theta _{\alpha, \beta} $. We use the shorthand  $s_{ij}\equiv \sin \theta _{ij}$ and
$c_{ij}\equiv \cos \theta _{ij}$ to refer to the mixing angles. In our numerical scans of the  parameter space in the model, we  fit to the best-fit experimental values  for the mixing angles:   $%
s_{13}^{2}=0.025_{-0.003}^{+0.003}$, $%
s_{12}^{2}=0.320_{-0.017}^{+0.016}$, $s_{23}^{2}=0.43_{-0.03}^{+0.03}$, and the mass-squared differences:  $\Delta
m_{21}^{2}=7.62_{-0.19}^{+0.19}\times 10^{-5}\mathrm{eV}^{2}$ and $|\Delta
m_{13}^{2}|=2.55_{-0.09}^{+0.06}\times 10^{-3}\mathrm{eV}^{2}$~\cite{Tortola:2012te}.

To determine the parameter space that generates viable neutrino masses, we
use the Casas-Ibarra parameterization~\cite{Casas:2001sr}
\begin{equation}
(\mathcal{M}_{\nu })_{\alpha \beta }=\sum_{i}g_{i\alpha }g_{i\beta }\Lambda
_{i}=\left( g^{T}\Lambda g\right) _{\alpha \beta },
\end{equation}%
with%
\begin{equation}
\Lambda _{i}=\frac{M_{i}}{16\pi ^{2}}\left\{ \frac{M_{S^{0}}^{2}}{%
M_{S^{0}}^{2}-M_{i}^{2}}\ln \frac{M_{S^{0}}^{2}}{M_{i}^{2}}-\frac{M_{A}^{2}}{%
M_{A}^{2}-M_{i}^{2}}\ln \frac{M_{A}^{2}}{M_{i}^{2}}\right\} .
\end{equation}%
According to the Casas-Ibarra parameterization, the coupling $g$ can be
written as%
\begin{equation}
g=D_{\sqrt{\Lambda ^{-1}}}\mathcal{R}D_{\sqrt{m_{\nu }}}U_{\nu }^{\dag },
\end{equation}
 where $D_{\sqrt{\Lambda ^{-1}}}=\mathrm{diag}\left\{ \sqrt{\Lambda _{1}^{-1}},%
\sqrt{\Lambda _{2}^{-1}},\sqrt{\Lambda _{3}^{-1}}\right\} $, $D_{\sqrt{%
m_{\nu }}}=\mathrm{diag}\left\{ \sqrt{m_{1}},\sqrt{m_{2}},\sqrt{m_{3}}\right\}$,  and $\mathcal{R}$\ is an orthogonal
rotation matrix ($m_{1,2,3}$ are the neutrino eigen-masses).
\section{Invisible Higgs Decays\label{sec:higgsdecay}}

The model is subject to constraints on the branching fraction for invisible Higgs decays, $%
\mathcal{B}(h\rightarrow inv)<17\%$~\cite{Hinv}. One should use $%
inv\equiv \{h_{2}h_{2}\},\{N_{\text{{\tiny DM}}}N_{\text{{\tiny DM}}}\}$,
when kinematically available, with corresponding decay widths given by%
\begin{align}
\Gamma \left( h_1\rightarrow h_2h_2\right) & =\frac{1}{32\pi }\frac{%
\left( \lambda_{122}\right)^{2}}{M_{h_{1}}}\,\left( 1-\frac{4M_{h_2}^{2}%
}{M_{h_1}^{2}}\right)^{\frac{1}{2}}\Theta \left(
M_{h_1}-2M_{h_2}\right), \notag \\
\Gamma \left( h_{1}\rightarrow N_{\text{{\tiny DM}}}N_{\text{{\tiny DM}}%
}\right) & =\frac{\tilde{y}_{\text{{\tiny DM}}}^{2}s_{h}^{2}}{16\pi }%
M_{h_{1}}\left( 1-\frac{4M_\dm^{2}}{M_{h_{1}}^{2}}\right) ^{%
\frac{3}{2}}\Theta \left( M_{h_1}-2M_\dm\right).
\end{align}%
The effective cubic coupling $\lambda_{122}$ is defined in Eq.~\eqref{eq:eff_couplings} below. As a  result of the SI symmetry, the coupling $\lambda_{122}$ vanishes at tree-level, and the non-zero loop-level coupling is
sufficiently small to ensure that decay to $h_{2}$ pairs is highly suppressed.%
\footnote{Note that $h_{2}$ decays to SM states, similar to a light SM Higgs boson but
with suppression by the mixing angle, $s_{h}^{2}$. However,  dedicated ATLAS or CMS searches for such light scalars, in the
channels $2b$, $2\tau $ or $2\gamma $, do not currently exist, so we classify the decay $%
h_{1}\rightarrow h_{2}h_{2}$ as invisible. In practice, however, the suppression of $%
\Gamma (h_{1}\rightarrow h_{2}h_{2})$ due to SI symmetry renders this point
moot.}

\section{Lepton Flavor Violating Decays\label{sec:constraints}}

The new fields give rise to one-loop contributions to $\mu\rightarrow e+\gamma$. Normalized relative to $\mathrm{Br}(\mu\rightarrow e\nu_\mu\bar{\nu}_e)$, the corresponding branching fraction is
\bea
\frac{\mathrm{Br}(\mu\rightarrow e \gamma)}{\mathrm{Br}(\mu\rightarrow e\nu_\mu\bar{\nu}_e)}=\frac{3(4\pi)^3\alpha_{em}}{4 \mathrm{G}_F^2} \left|A_D\right|^2,
\eea
where $A_D$ is the dipole form factor:
\bea
A_D=\sum_i \frac{g_{ei}^* g_{i\mu}}{32\pi^2}\frac{1}{M_{S^+}^2}\,F^{(n)}(M_i^2/M_{S^+}^2).
\eea
with the loop function given by 
\bea
F^{(n)}(x)=[1-6x+3x^2+2x^3-6x^2\ln x]/[6(1-x)^4].  
\eea
A simple change of labels allows one to use the above formulae for the related decay $\tau\rightarrow \mu+\gamma$. In our analysis we also include the constraint from neutrino-less double beta decay.

Note that, in general, the scotogenic model is subject to strong LFV constraints, relating to the fact that the DM annihilates via the same Yukawa couplings that mediate LFV processes. Consequently one cannot decouple the two effects and there can be tension between the demands of suppressed LFV processes and the attainment of a viable DM abundance (actually, in the scotogenic model, constraints from other LFV processes, like $\mu$-$e$ conversion, can be more severe than the above LFV decays; see the 3rd and 4th papers in Ref.~\cite{Schmidt:2012yg}). However, we shall see that the situation differs in the SI model, due to additional annihilation processes mediated by the dilaton. This provides a degree of decoupling between the LFV processes and DM annihilations, such that LFV bounds are more readily satisfied. Thus, for our purposes, it is sufficient to consider the above LFV decays (we shall see that the viable parameter space includes regions well-below the LFV bounds, so slightly stronger bounds do not have a large effect). We note that the correlation between $\mu\rightarrow e\gamma$ and the DM relic abundance, for the case of fermionic DM in the scotogenic model, was first noted in Ref.~\cite{Kubo:2006yx}, while Ref.~\cite{Babu:2007sm} noted that models with a singlet scalar allow one to decouple these issues.

\section{Dark Matter\label{sec:dark_matter}}

\subsection{Relic Density}

As the universe cools, the temperature eventually drops below the DM
mass. Consequently the  DM number density becomes Boltzmann  suppressed and  the DM annihilation
rate can become comparable to the Hubble parameter. At a
certain temperature the
DM particles freeze out of equilibrium, such that the DM number density
in a comoving volume henceforth remains constant.  The cold DM\ relic
abundance therefore depends on the total thermally averaged annihilation
cross section%
\begin{eqnarray}
& &\left\langle \sigma (N_{\text{{\tiny DM}}}\ N_{\text{{\tiny DM}}%
})v_{r}\right\rangle=\sum_{X}\left\langle \sigma (N_{\text{{\tiny DM}}}\
N_{\text{{\tiny DM}}}\rightarrow X)v_{r}\right\rangle  \notag \\
&=& \sum_{X}\int_{4M_\dm^{2}}^{\infty }ds~\sigma _{N_{\text{%
{\tiny DM}}}\ N_{\text{{\tiny DM}}}\rightarrow X}(s)\frac{\left( s-4M_{\text{%
{\tiny DM}}}^{2}\right)}{8TM_{\text{%
{\tiny DM}}}^{4}K_{2}^{2}\left( \frac{M_\dm}{T}\right) } \sqrt{s}K_{1}\left( \frac{\sqrt{s}}{T}\right) ,
\end{eqnarray}%
where $v_{r}$ is the relative velocity, $s$ is the Mandelstam variable, $%
K_{1,2}$ are the modified Bessel functions and $\sigma _{N_{\text{{\tiny DM}}%
}\ N_{\text{{\tiny DM}}}\rightarrow X}(s)$ is the annihilation cross due to
the channel $N_{\text{{\tiny DM}}}\ N_{\text{{\tiny DM}}}\rightarrow X$,  at
the CM energy $\sqrt{s}$. At freeze-out, the thermal relic density can
be given in terms of the thermally averaged annihilation cross section by
\begin{equation}
\Omega _{\text{{\tiny DM}}}h^{2}\simeq \frac{(1.07\times 10^{9})x_{F}}{\sqrt{%
g_{\ast }}M_{pl}(\mathrm{GeV})\left\langle \sigma (N_{\text{{\tiny DM}}}\ N_{%
\text{{\tiny DM}}})v_{r}\right\rangle },
\end{equation}%
where $M_{pl}$ is the Plank mass and $g_{\ast }$ counts the effective degrees
of freedom of the relativistic fields in equilibrium. The inverse
freeze-out temperature, $x_{F}=M_\dm/T_{F}$, can be determined
iteratively from the equation%
\begin{equation}
x_{F}=\log \left( \sqrt{\frac{45}{8}}\frac{M_{\text{{\tiny DM}}%
}M_{pl}\left\langle \sigma (N_{\text{{\tiny DM}}}N_{\text{{\tiny DM}}%
})v_{r}\right\rangle }{\pi ^{3}\sqrt{g_{\ast }x_{F}}}\right) .
\end{equation}

\begin{figure}[h]
\begin{center}
\includegraphics[width = 0.8\textwidth]{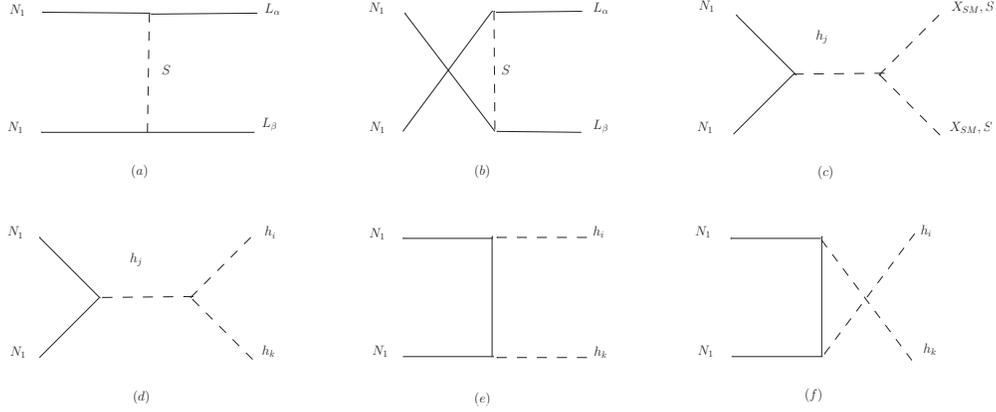}
\end{center}
\caption{Diagrams for DM annihilation.}
\label{DM-ahn}
\end{figure}

In the present model, the  classes of DM annihilation channels are shown in
Fig.~\ref{DM-ahn}. The DM can annihilate into: (1)   charged leptons and neutrinos, $\ell
_{\alpha }^{-}\ell _{\beta }^{+}$ and $\nu _{\alpha }\bar{\nu}_{\beta }$, including LFV final states with $\alpha\ne\beta$,\
(2) SM fermions and gauge bosons $b\bar{b}$, $t\bar{t}$, $W^{+}W^{-}$, $ZZ$
and the scalars $SS$, and (3) final states comprised of the Higgs and/or dilaton, $%
h_{i}h_{k}$. The first class of channels are $h_{1,2}$-mediated $s$-channel
processes, the second class are $S$-mediated $t$-channel processes
while the third class contains both $s$- and $t$-channels
processes mediated by $h_{1,2}$.


\subsection{Annihilation Cross Sections}

\textbf{(1) t-channel processes}

The cross section for the annihilation channel into charged leptons\footnote{For same-flavor charged leptons ($\alpha=\beta $), there are also  $s$-channel processes mediated by $h_{1,2} $. However, these are proportional to their
Yukawa couplings and may therefore be ignored.} is given by~\cite{Cheung:2004xm}%
\begin{eqnarray}
\sigma (N_{\text{{\tiny DM}}}N_{\text{{\tiny DM}}} &\rightarrow &\ell
_{\alpha }^{-}\ell _{\beta }^{+})v_{r}=\frac{1}{8\pi }\frac{|g_{1\alpha
}g_{1\beta }^{\ast }|^{2}}{s(M_{S^{+}}^{2}-M_\dm^{2}+\frac{s%
}{2})^{2}}\left[ \frac{m_{\ell _{\alpha }}^{2}+m_{\ell _{\beta }}^{2}}{2}%
\left( \frac{s}{2}-M_\dm^{2}\right) \right.  \notag \\
&&\left[ +\frac{2}{3}s\left( \frac{s}{4}-M_\dm^{2}\right)
\frac{(M_{S^{+}}^{2}-M_\dm^{2})^{2}+\frac{s}{2}%
(M_{S^{+}}^{2}-M_\dm^{2})+\frac{s^{2}}{8}}{(M_{S^{+}}^{2}-M_{%
\text{{\tiny DM}}}^{2}+\frac{s}{2})^{2}}\right] .  \label{Sll}
\end{eqnarray}%
The cross section for annihilation  into neutrinos can be obtained from Eq.~(\ref%
{Sll}) by replacing $M_{S^{+}}^{2}\rightarrow M_{S^{0}}^{2}$ and sending the
charged lepton masses to zero, i.e.,%
\begin{equation}
\sigma (N_{\text{{\tiny DM}}}N_{\text{{\tiny DM}}}\rightarrow \nu _{\alpha
}\nu _{\beta })v_{r}=\frac{|g_{1\alpha }g_{1\beta }^{\ast }|^{2}}{12\pi }%
\left( \frac{s}{4}-M_\dm^{2}\right) \frac{(M_{S^{0}}^{2}-M_{%
\text{{\tiny DM}}}^{2})^{2}+\frac{s}{2}(M_{S^{0}}^{2}-M_{\text{{\tiny DM}}%
}^{2})+\frac{s^{2}}{8}}{(M_{S^{0}}^{2}-M_\dm^{2}+\frac{s}{2}%
)^{4}}.
\end{equation}

\textbf{(2) s-channel processes}

The processes $N_{\text{{\tiny DM}}}N_{\text{{\tiny DM}}}\rightarrow b\bar{b}
$, $t\bar{t}$, $W^{+}W^{-}$ and $ZZ$ can occur as shown in Fig. \ref{DM-ahn}%
-c. The corresponding amplitude can be written as%
\begin{equation}
\mathcal{M}=ic_{h}s_{h}y_{1}\bar{u}\left( k_{2}\right) u\left( k_{1}\right)
\left( \frac{i}{s-M_{h_{1}}^{2}}-\frac{i}{s-M_{h_{2}}^{2}}\right) \mathcal{M}%
_{h\rightarrow SM}\left( m_{h}\rightarrow \sqrt{s}\right) ,
\end{equation}%
with $\mathcal{M}_{h\rightarrow SM}\left( m_{h}\rightarrow \sqrt{s}\right) $
being the amplitude of the Higgs decay $h\rightarrow X_{SM}\bar{X}_{SM}$, with
the Higgs mass replaced as $m_{h}\rightarrow \sqrt{s}$. This leads to the
cross section%
\begin{equation}
\sigma (N_{\text{{\tiny DM}}}N_{\text{{\tiny DM}}}\rightarrow X_{SM}\bar{X}%
_{SM})\upsilon _{r}=8\sqrt{s}s_{h}^{2}c_{h}^{2}y_{1}^{2}\left\vert \frac{1}{%
s-M_{h_{1}}^{2}}-\frac{1}{s-M_{h_{2}}^{2}}\right\vert ^{2}\Gamma
_{h\rightarrow X_{SM}\bar{X}_{SM}}\left( m_{h}\rightarrow \sqrt{s}\right) ,
\label{csSM}
\end{equation}%
where $\Gamma _{h\rightarrow X_{SM}\bar{X}_{SM}}\left( m_{h}\rightarrow
\sqrt{s}\right) $\ is the total decay width, with $m_{h}\rightarrow \sqrt{s}$.

Similarly, the $SS$ annihilation cross section can written as%
\begin{equation}
\sigma (N_{\text{{\tiny DM}}}N_{\text{{\tiny DM}}}\rightarrow SS)v_{r}=\eta
_{S}\frac{s_{h}^{2}c_{h}^{2}y_{1}^{2}}{4\pi }s\left\vert \frac{c_{h}\lambda
_{1SS}}{s-M_{h_{1}}^{2}-iM_{h_{1}}\Gamma _{h_{1}}}+\frac{s_{h}\lambda _{2SS}%
}{s-M_{h_{2}}^{2}-iM_{h_{2}}\Gamma _{h_{2}}}\right\vert ^{2}\left( 1-\frac{%
4M_{S}^{2}}{s}\right) ^{1/2}
\end{equation}%
where $\eta _{S^{0}}=\eta _{A}=1,~\eta _{S^{+}}=2$, and $\lambda _{1SS}$\
and\ $\lambda _{2SS}$\ are the triple couplings of a scalar $h_{1,2}$
with two $S$ fields, given by%
\begin{eqnarray}
\lambda _{1S^{+}S^{-}} &=&\lambda _{3}c_{h}v-\lambda _{\phi \text{{\tiny S}}%
}s_{h}x,~\lambda _{2S^{+}S^{-}}=\lambda _{3}s_{h}v+\lambda _{\phi \text{%
{\tiny S}}}c_{h}x,  \notag \\
\lambda _{1S^{0}S^{0},1AA} &=&\frac{1}{2}\left( \lambda _{3}+\lambda _{4}\pm
\lambda _{5}\right) c_{h}v-\frac{1}{2}\lambda _{\phi \text{{\tiny S}}}s_{h}x,
\notag \\
\lambda _{2S^{0}S^{0},2AA} &=&\frac{1}{2}\left( \lambda _{3}+\lambda _{4}\pm
\lambda _{5}\right) s_{h}v+\frac{1}{2}\lambda _{\phi \text{{\tiny S}}}c_{h}x.
\end{eqnarray}

\textbf{(3) Higgs channel}

The DM can self-annihilate into $h_{i}h_{k}$,\ as seen in Fig. \ref{DM-ahn}%
-d, -e and -f. The amplitude squared is given by%
\begin{eqnarray}
\left\vert \mathcal{M}\right\vert ^{2} &=&2\tilde{y}_{\text{{\tiny DM}}}^{2}s%
\left[ {\frac{c_{{h}}\lambda _{{1ik}}}{s-{M_{h_{1}}^{2}}}}+{\frac{s_{{h}%
}\lambda _{{2ik}}}{s-{M_{h_{2}}^{2}}}}\right] ^{2}  \notag \\
&&+4c_{{i}}c_{{k}}\tilde{y}_{\text{{\tiny DM}}}^{3}M_\dm%
\left[ {\frac{c_{{h}}\lambda _{{1ik}}}{s-{M_{h_{1}}^{2}}}}+{\frac{s_{{h}%
}\lambda _{{2ik}}}{s-{M_{h_{2}}^{2}}}}\right] \left( \frac{s-{M_{h_{i}}^{2}}+%
{M_{h_{k}}^{2}}}{t-M_\dm^{2}}+a\frac{s+{M_{h_{i}}^{2}}-{%
M_{h_{k}}^{2}}}{u-M_\dm^{2}}\right)  \notag \\
&&+\frac{2c_{{i}}^{2}c_{{k}}^{2}\tilde{y}_{\text{{\tiny DM}}}^{4}}{\left(
t-M_\dm^{2}\right) ^{2}}\left\{ 4M_\dm^{2}{%
M_{h_{k}}^{2}}+\left( M_\dm^{2}+{M_{h_{i}}^{2}}-t\right)
\left( M_\dm^{2}+{M_{h_{i}}^{2}}-u\right) -s{M_{h_{i}}^{2}}%
\right\}  \notag \\
&&+a^{2}\frac{2c_{{i}}^{2}c_{{k}}^{2}\tilde{y}_{\text{{\tiny DM}}}^{4}}{%
\left( u-M_\dm^{2}\right) ^{2}}\left\{ 4M_{\text{{\tiny DM}}%
}^{2}{M_{h_{i}}^{2}}+\left( M_\dm^{2}+{M_{h_{k}}^{2}}%
-u\right) \left( M_\dm^{2}+{M_{h_{k}}^{2}}-t\right) -s{%
M_{h_{k}}^{2}}\right\}  \notag \\
&&+a\frac{2c_{{i}}^{2}c_{{k}}^{2}\tilde{y}_{\text{{\tiny DM}}}^{4}}{\left(
t-M_\dm^{2}\right) \left( u-M_\dm^{2}\right)
}\left\{ \left( M_\dm^{2}+{M_{h_{i}}^{2}}-t\right) \left( M_{%
\text{{\tiny DM}}}^{2}+{M_{h_{k}}^{2}}-t\right) \right.  \notag \\
&&\left. +\left( M_\dm^{2}+{M_{h_{k}}^{2}}-u\right) \left(
M_\dm^{2}+{M_{h_{i}}^{2}}-u\right) -\left( s-4M_{\text{%
{\tiny DM}}}^{2}\right) \left( s-M_{h_{i}}^{2}-M_{h_{k}}^{2}\right) \right\}
,
\end{eqnarray}%
with $s$, $t$ and $u$ being the Mandelstam variables, and the Yukawa couplings
are defined as $\tilde{y}_{\text{{\tiny DM}}}\equiv y_{1}$, $c_{1}\equiv c_{h}$
and $c_{2}\equiv s_{h}$. Here, we integrate the phase space numerically to obtain the cross section for a given value of $s$. At tree-level the
effective cubic scalar couplings ($\lambda _{1ik}$ and $\lambda _{2ik})$ are
given by
\begin{eqnarray}
\lambda _{111} &=&6\lambda _{\text{{\tiny H}}}\ c_{h}^{3}v-3\lambda _{\phi
\text{{\tiny H}}}c_{h}^{2}s_{h}v+3\lambda _{\phi \text{{\tiny H}}%
}c_{h}s_{h}^{2}x-6\lambda _{\phi }s_{h}^{3}x,  \notag \\
\lambda _{112} &=&\lambda _{\phi \text{{\tiny H}}%
}c_{h}^{3}x+2c_{h}^{2}s_{h}(3\lambda _{\text{{\tiny H}}}-\lambda _{\phi
\text{{\tiny H}}})v+2c_{h}s_{h}^{2}(3\lambda _{\phi }-\lambda _{\phi \text{%
{\tiny H}}})x+\lambda _{\phi \text{{\tiny H}}}s_{h}^{3}v,  \notag \\
\lambda _{222} &=&\lambda _{122}\ =\ 0,  \label{eq:eff_couplings}
\end{eqnarray}%
though for completeness we employ the one-loop results, obtained
from the loop-corrected potential following Ref.~\cite{AAN}. We note that the (leading order) absence of the cubic
interactions $h_{1}h_{2}^{2}$ and $h_{2}^{3}$, is a
general feature of SI models.

\subsection{Direct Detection}

With regard to  direct-detection experiments, interactions between the DM and quarks are described by an effective low-energy Lagrangian:
\begin{equation}
\mathcal{L}_{N_{1}-q}^{(eff)}=a_{q}\,\bar{q}q\,N_{\text{{\tiny DM}}%
}^{c}N_{\text{{\tiny DM}}},
\end{equation}%
with%
\begin{equation}
a_{q}=-\frac{s_{h}c_{h}M_{q}M_\dm}{2\left\langle \phi \right\rangle \langle H^{0}\rangle}\left[ \frac{1}{M_{h_{1}}^{2}}-\frac{1}{%
M_{h_{2}}^{2}}\right] .
\end{equation}%
Consequently, the effective nucleon-DM interaction is written as%
\begin{equation*}
\mathcal{L}_{\text{{\tiny DM}}-\mathcal{N}}^{(eff)}=a_{\mathcal{N}}\mathcal{%
\bar{N}N}N_{\text{{\tiny DM}}}^{c}N_{\text{{\tiny DM}}},
\end{equation*}%
where
\begin{equation}
a_{\mathcal{N}}=\frac{s_{h}c_{h}\left( M_{\mathcal{N}}-\frac{7}{9}M_{%
\mathcal{B}}\right) M_\dm}{\left\langle
\phi \right\rangle \langle H^{0}\rangle  }\left[ \frac{1}{M_{h_{1}}^{2}}-\frac{1}{M_{h_{2}}^{2}}%
\right] .
\end{equation}%
In this relation, $M_{\mathcal{N}}$ is the nucleon mass and $M_{\mathcal{B}}$
the baryon mass in the chiral limit \cite{He}. This leads to the following
nucleon-DM elastic cross section in the chiral limit%
\begin{equation}
\sigma _{\det }=\frac{s_{h}^{4}M_{\mathcal{N}}^{2}\left( M_{\mathcal{N}}-%
\frac{7}{9}M_{\mathcal{B}}\right) ^{2}M_\dm^{4}}{\pi \langle
H^{0}\rangle ^{4}\left( M_\dm+M_{\mathcal{B}}\right) ^{2}}%
\left[ \frac{1}{M_{h_{1}}^{2}}-\frac{1}{M_{h_{2}}^{2}}\right] ^{2}.
\end{equation}%
The analysis below will show that the upper bound reported by LUX experiment \cite{Akerib:2013tjd} provides a stringent constraint on $\sigma _{det}$.

\section{Analysis and Results \label{sec:results}}

Next we turn to our numerical analysis and results. We perform a numerical scan of the parameter space to determine whether radiative electroweak symmetry breaking is compatible with one-loop radiative neutrino mass and singlet neutrino DM. In the scans, we enforce the minimization conditions, Eqs.~\eqref{1loopcoupling_condition} and
  \eqref{eq:1loopvev_cond}, vacuum stability via  Eq.~\eqref{1loopstability_condition}, and demand that the 
SM-like Higgs mass is in the experimentally allowed range, $M_{h_{1}}=125.09\mp 0.21$~\textrm{GeV}. Compatibility with
constraints from LEP (OPAL) on a light Higgs~\cite{OPAL} are enforced, and we consider the constraint
from the Higgs invisible decay, $\mathcal{B}(h\rightarrow inv)<17\%$, \cite%
{Hinv}.  Dimensionless couplings are restricted to the
perturbative range throughout,  and we consider values of $100~\mathrm{GeV}<\langle \phi
\rangle <5$~\textrm{TeV} for the beyond-SM VEV (however, we only find viable benchmark points for $\langle \phi\rangle \gtrsim150$~GeV).\footnote{In principle, one can consider larger values for $\langle \phi\rangle$. However, these require hierarchically small couplings in the scalar potential~\cite{Foot:2013hna}, which we do not consider here.}

\begin{figure}[t]
\begin{center}
\includegraphics[width = 0.6\textwidth]{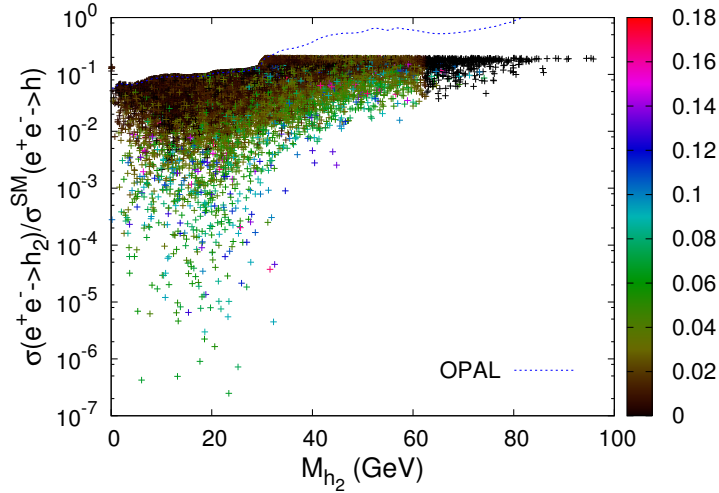}
\end{center}
\caption{Scalar mixing versus the light scalar mass $M_{h_2}$. The palette shows the
branching ratio for invisible Higgs decays. An overwhelming majority of
the points satisfy the constraint $B(h_{1}\rightarrow inv)<17\%$.}
\label{fig:mh2}
\end{figure}

The scan reveals a spread of viable
values for the dilaton mass $M_{h_{2}}$, consistent with  OPAL, as plotted in Figure~%
\ref{fig:mh2}. In the scan we tend to find $%
M_{h_{2}} $ in the range $\mathcal{O}(1)~\mathrm{GeV}\lesssim
M_{h_{2}}\lesssim90$~\textrm{GeV}. Lighter values of $M_{h_2}$
seemingly require an amount of engineered cancellation among the
radiative mass-corrections from fermions and bosons, or larger values for $\langle \phi\rangle$; see
Eq.~\eqref{eq:pgb_mass}. We noticed that regions with $\langle \phi
\rangle \gtrsim 500$~\textrm{GeV} tend to be preferred.

\begin{figure}[h]
\begin{center}
\includegraphics[width = 0.5\textwidth]{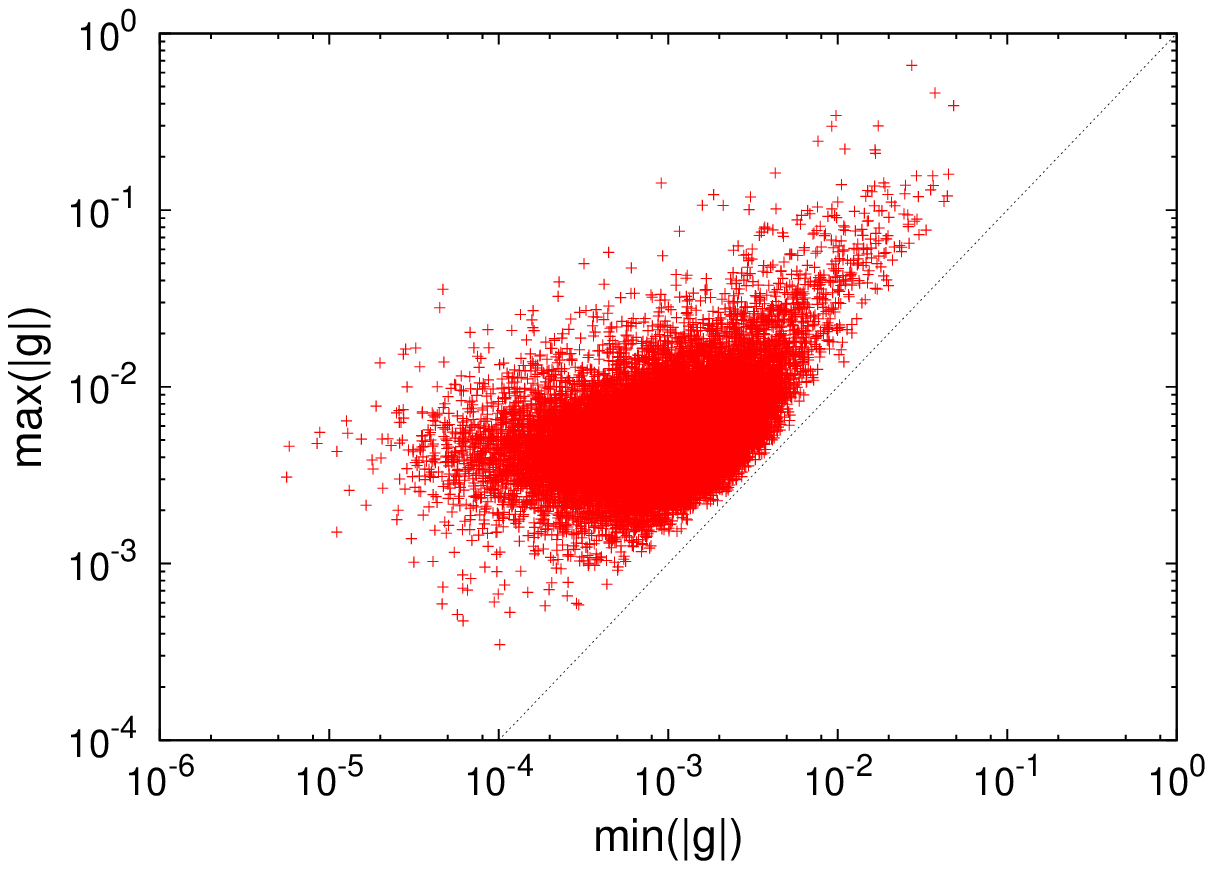}~%
\includegraphics[width =
0.5\textwidth]{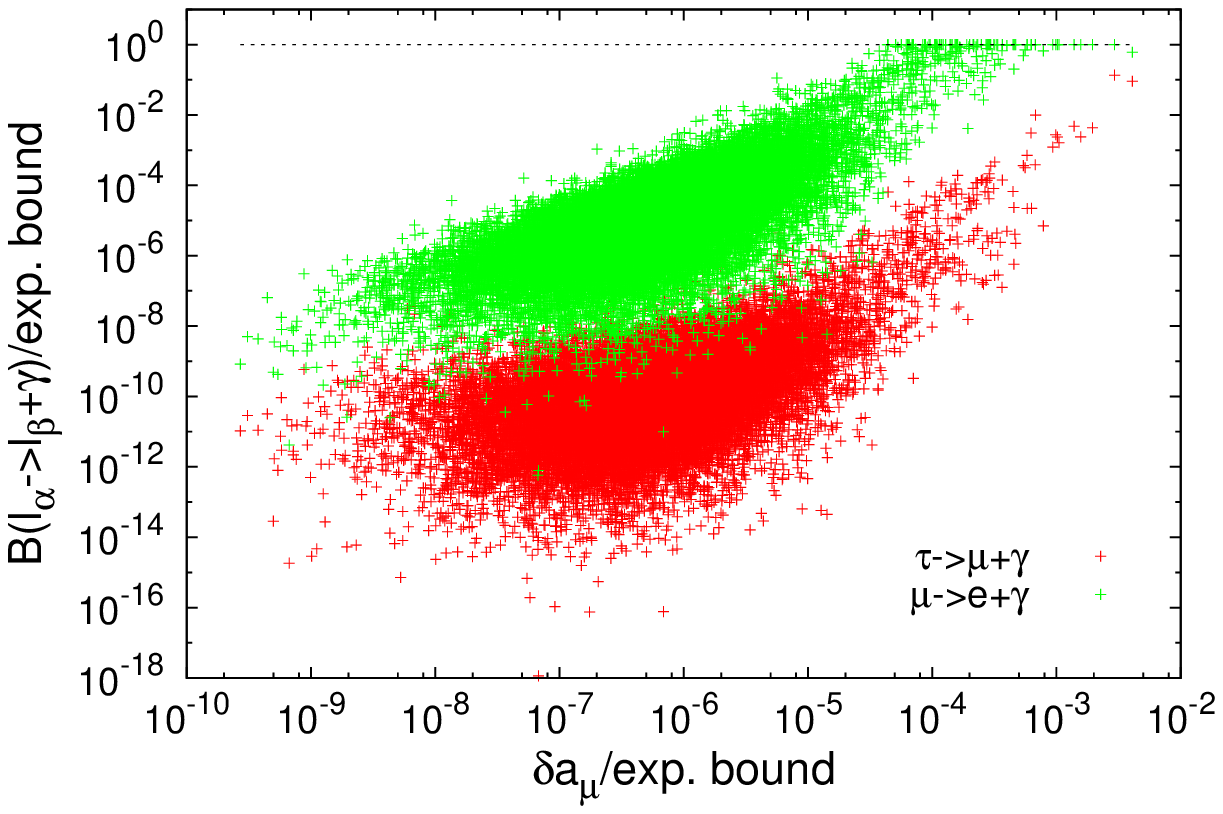}
\end{center}
\caption{Left: Viable benchmark points for the Yukawa couplings $g_{i\alpha}$, in absolute values. The
dashed line denotes the degenerate case, i.e, $\min \left\vert
g\right\vert =\max \left\vert g\right\vert $. Right: The LFV branching
ratios versus the muon anomalous
magnetic moment, both scaled by the experimental bounds. }
\label{fig:g_LFV}
\end{figure}

We further scan  for parameter space giving viable neutrino masses and mixing, subject to the LFV and
muon anomalous magnetic moment constraints, while simultaneously generating a viable DM
relic density. Figure~\ref{fig:g_LFV} shows viable benchmark sets for the
Yukawa couplings $g_{i\alpha}$, along with the
corresponding LFV branching ratios and $\delta a_\mu$ contributions. The couplings $g_{i\alpha}$ are typically well-below the perturbative bound.  Note that  the range for the Yukawa couplings varies over several orders of magnitude. This reflects the freedom to  take the lepton-number violating quartic coupling $\lambda_5$ to be small, and accordingly transfer some of the neutrino mass suppression between the Yukawa and quartic coupling sectors. The capacity to obtain viable neutrino masses, with Yukawa couplings that vary over a considerable range, influences the strength of the signal  from LFV decays.  Figure~\ref{fig:g_LFV} shows that the bound from $\mu\rightarrow e\gamma$ gives important constraints in parameter space with larger $g_{i\alpha}$, while smaller values of $g_{i\alpha}$ allow the model to easily evade the bound. Constraints from the weaker $\tau\rightarrow\mu\gamma$ bound
are readily satisfied.  Also, we verified that constraints from
neutrino-less double-beta decay searches are satisfied by the
benchmark points.

\begin{figure}[h]
\begin{center}
\includegraphics[width = 0.5\textwidth]{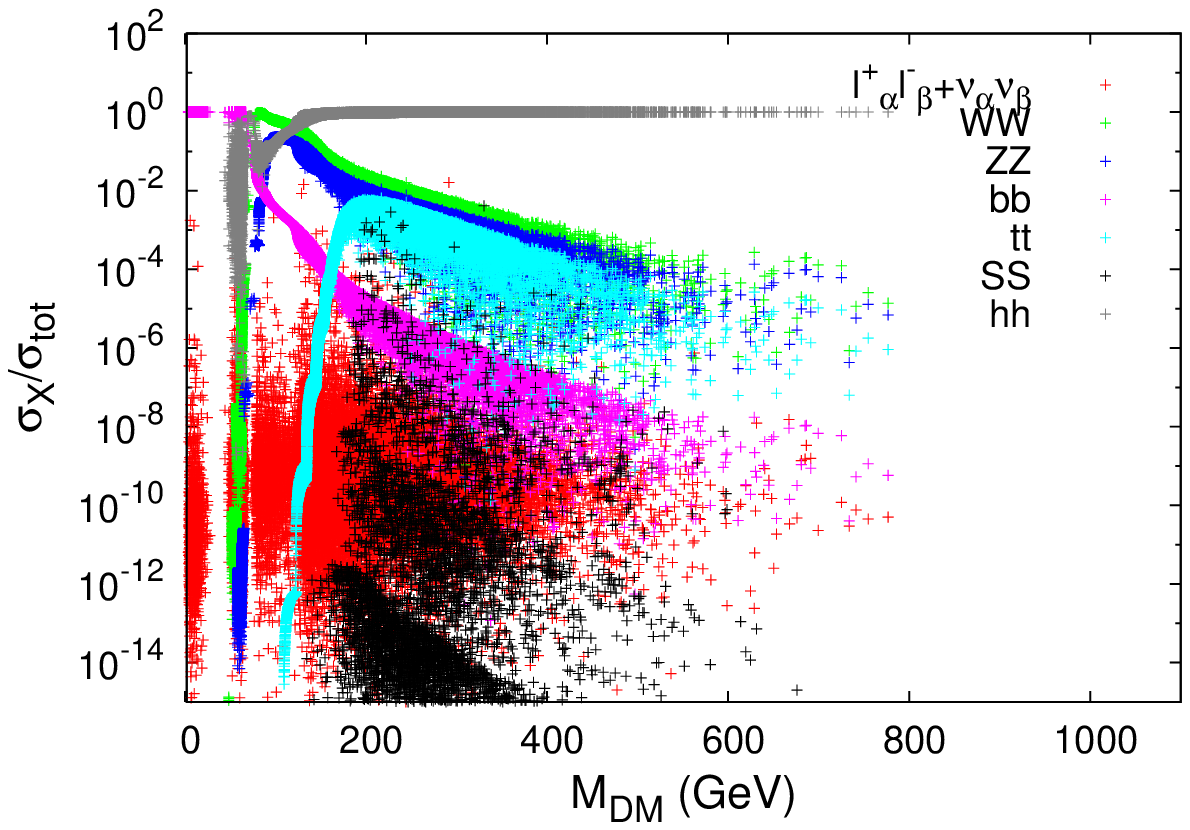}~\includegraphics[width=0.5%
\textwidth]{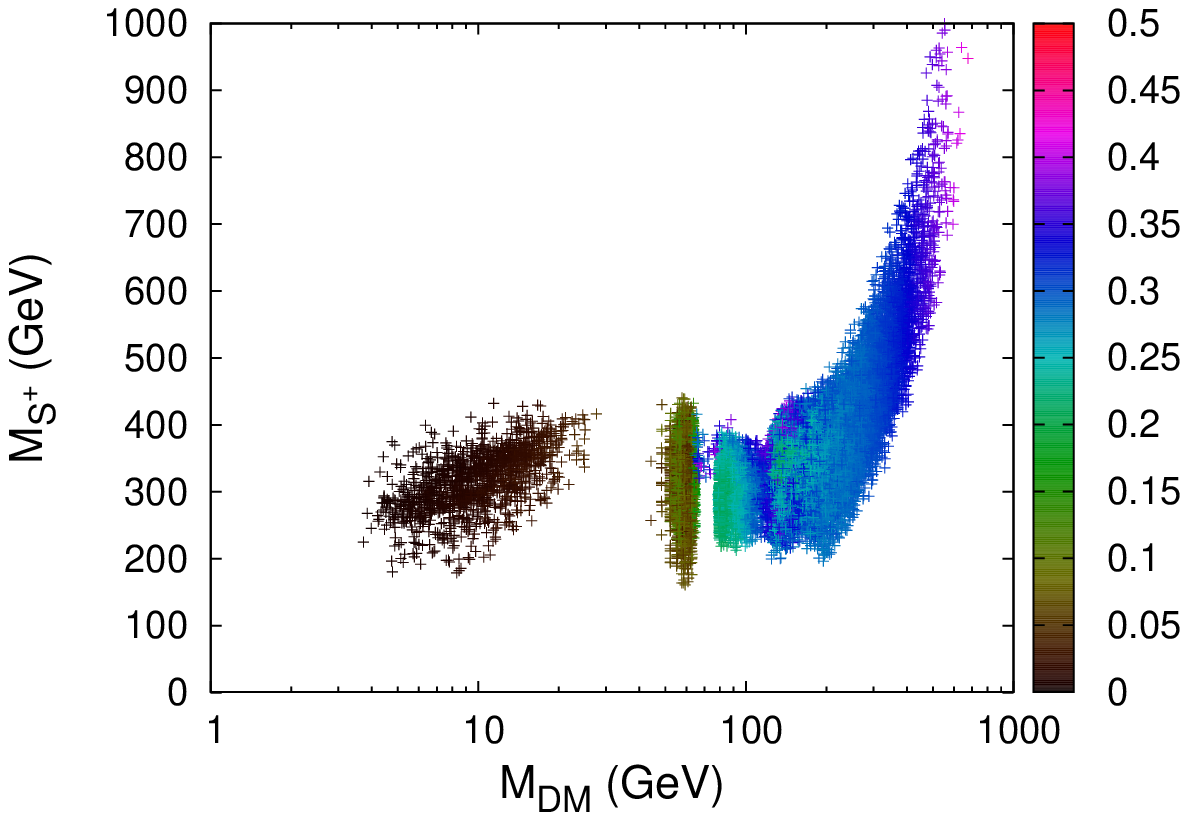}
\end{center}
\caption{Left: The cross section ratio $\protect\sigma _{X}/\protect\sigma_{\mathrm{tot}}$\ at freeze-out versus the DM mass. Here $X$ denotes lepton pairs, gauge bosons, heavy quarks and scalars.  Right: The
charged scalar masses $M_{S^+}$ versus the DM mass. The palette shows the DM Yukawa coupling $y_\dm\equiv y_1$.}
\label{DM}
\end{figure}

With regards to the DM relic density, recall that there are multiple classes of
annihilation channels, namely $N_{\text{{\tiny DM}}}N_{\text{{\tiny DM}}%
}\rightarrow X$ ($X=\ell_{\alpha}^{\mp}\ell_{\beta}^{\pm}$, $\nu_\alpha\nu_\beta$,
$b\bar{b}$, $t\bar{t}$, $WW$, $ZZ$, $SS$, $h_{1,2}h_{1,2}$). Depending on the specific value of the
DM mass, a given channel may be significant or suppressed. To probe the role of the distinct channels, in Figure~\ref{DM}-left we plot the   contribution
of each channel relative  to the total cross section at freeze-out, 
$\sigma_{X}/\sigma_{tot}$, versus the DM
mass.  Annihilations into lepton pairs typically play a subdominant role. These are mediated by the couplings $g_{i\alpha}$, whose values should be sufficiently small to ensure viable neutrino masses and consistency with LFV constraints. For lighter values of $M_{\dm}\lesssim 75$~GeV,  the cross section tends to be dominated by annihilations into $b$ quarks, while annihilations into $Z_2$-even neutral scalar final states ($X=hh$ with $h\equiv h_{1,2}$) are dominant for heavier values of $M_\dm\gtrsim125$~GeV. In the intermediate
range, annihilations into gauge bosons can also be important.  For completeness, we include the final states $X=2S$ in the plot, for components of the doublet $S$.  Although the doublet scalars are typically heavier than the DM, thermal fluctuations can allow a contribution from these modes (though the effect is clearly subdominant, as seen in the Figure). Figure~\ref{DM}-right shows the
 mass  of the charged scalar, $M_{S^+}$, versus the DM mass. In the lighter DM mass range, $M_\dm\lesssim\mathcal{O}(100)$~GeV, one notices that the charged scalar mass should not exceed 450~GeV, while for larger values of $M_\dm$ one can have $M_{S^+}$  at the TeV
scale. Such light charged scalars may be of phenomenological interest as they can be within reach of collider
experiments. 

\begin{figure}[h]
\begin{center}
\includegraphics[width = 0.6\textwidth]{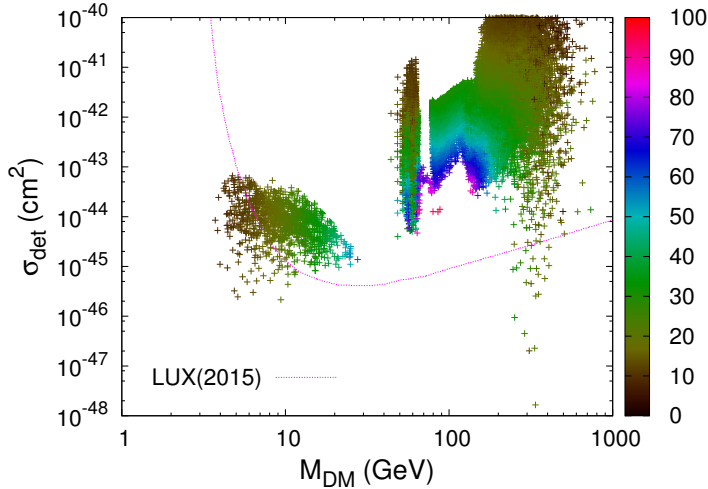}
\end{center}
\caption{The direct detection cross section versus the DM mass. The dashed line shows the
the recent constraints from LUX, while the palette gives the mass for the neutral
beyond-SM scalar (dilaton), $M_{h_2}$, in units of GeV.}
\label{fig:DDet}
\end{figure}

We note that Figure~\ref{DM}-right contains disconnected regions for viable DM, with the region $31~\mathrm{GeV}\lesssim M_\dm\lesssim 48$~GeV not returning viable benchmark points. This ``missing region" results from an over-abundance of DM, due to an insufficiently large, thermally-averaged  annihilation cross section. In the small $M_{\dm}$ region, the annihilation cross section is dominated by $b\bar{b}$ final states, with an important sub-contribution from annihilations into  dilatons. However, below $M_\dm\approx 48$~GeV,  we find that the dilaton contribution is too small to  allow the observed relic abundance. The allowed island at $M_\dm\lesssim 31$~GeV corresponds to parameter space that approaches the $h_2$ resonance, such that $2M_\dm$ is around, or just below, the dilaton mass, namely $M_\dm\lesssim M_{h_2}/2$ (the dilaton mass is shown in Figure~\ref{fig:DDet}). This enhances annihilations into SM final states. The corresponding enhancement to the $s$-channel process $N_\dm N_\dm\rightarrow h_2h_2$, via an intermediate $h_2$, is not sufficient to overcome the small cubic coupling $\lambda_{222}$, as shown in Eq.~\eqref{eq:eff_couplings}. Note also that points in the region $M_\dm\lesssim M_{h_1}/2\approx 60$~GeV experience some enhancement from the $h_1$ resonance. Such enhancements do not occur in heavier $M_\dm$ regions, as both the dilaton and Higgs are much lighter than the DM. Throughout the lighter $M_\dm$ regions, the Higgs may decay into $N_\dm$ and $h_2$ final states, though the bound on invisible Higgs decays is readily satisfied. The decay $h_1\rightarrow N_\dm N_\dm$ is sufficiently small due to Yukawa suppression (in addition to small $\theta_h$ mixing), as seen from the palette in Figure~\ref{DM}-right, while the decay $h_1\rightarrow h_2h_2$ is suppressed by the small cubic scalar coupling $\lambda_{122}$.

Next we consider the constraints from direct-detection experiments. We plot the direct-detection cross section
 versus  the DM mass for the benchmark parameter sets in Figure~\ref{fig:DDet}. The mass of the dilaton, $M_{h_2}$, in units of GeV, is shown in the corresponding palette. One immediately observes that 
direct-detection limits from LUX~\cite{Akerib:2013tjd} impose very serious constraints on the model, with a
large number of  benchmark sets already excluded. The plot shows that the surviving 
 benchmark points mostly occur for $M_\dm\lesssim 10$~%
\textrm{GeV}, with a smaller number of viable points found for $M_\dm\gtrsim 200$~\textrm{GeV}. Benchmarks with
intermediate $M_\dm$ values are excluded. The viable parameter space typically requires a lighter dilaton mass, $M_{h_2}\lesssim 10$~GeV, as all benchmarks with $M_{h_2}\gtrsim 50$~GeV  are excluded. It is clear from the figure that the surviving benchmark sets can be
 probed in forthcoming direct-detection experiments. 

\begin{figure}[h]
\begin{center}
\includegraphics[width =0.5\textwidth]{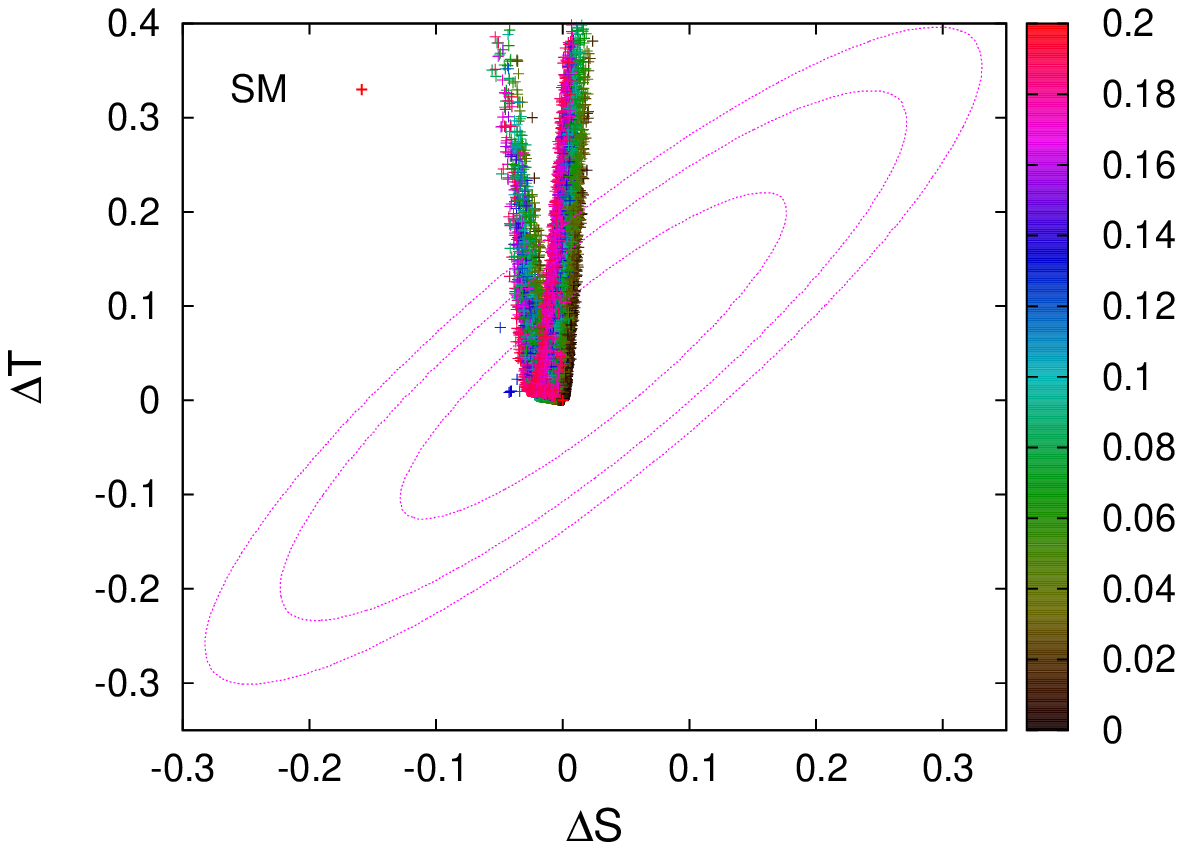}~\includegraphics[width=0.5%
\textwidth]{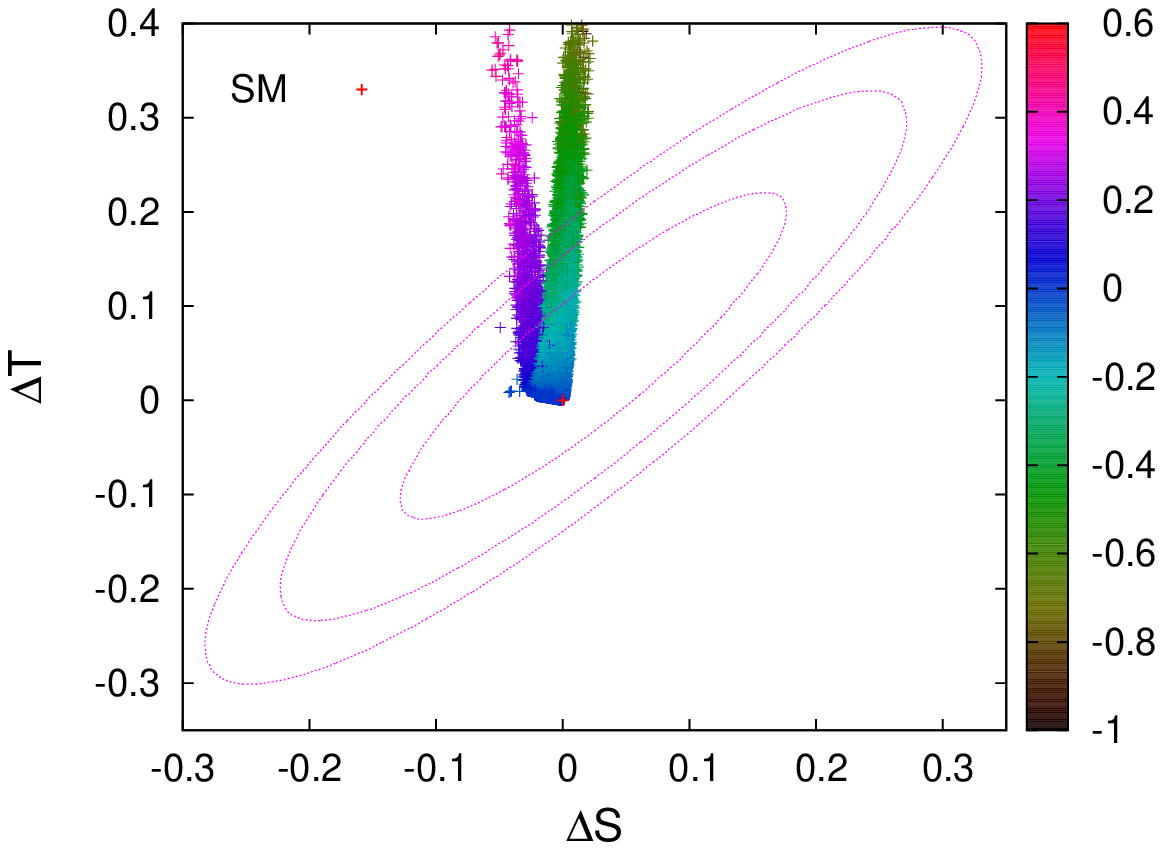}
\end{center}
\caption{Left: The oblique parameters $\Delta S$\ versus $\Delta T$\
for the benchmarks used previously. The ellipsoids show the
68\%, 95\% and 99\% CL., respectively. In the Left frame, the palette shows the mixing
$\sin ^{2}\theta_{h}$ between the Higgs and the dilaton; in the Right frame it shows the relative mass splitting, $\Delta =\left(2M_{S^+}- M_A-M_{S^0}\right)/2M_{S^+}$, for components of the scalar doublet $S$.} \label{fig:obl}
\end{figure}

In Figure~\ref{fig:obl} we consider the oblique parameters. The variation with respect to the mixing parameter $\sin^2\theta_h$ is shown in the left panel. One notices that the $\sin^2\theta_h$ dependence is not the dominant source of variation.  There is some sensitivity to  $\sin^2\theta_h$, primarily in $\Delta S$. However, for a given fixed value of $\sin^2\theta_h$, benchmark points occur along the majority of the V-shaped curve traced out in the plot. Thus, the $\sin^2\theta_h$ dependence is not driving the variation. The dependence of the oblique parameters on the dimensionless mass-difference for components of  $S$, namely $\Delta =\left(2M_{S^+}- M_A-M_{S^0}\right)/2M_{S^+}$, is shown in the right panel of Figure~\ref{fig:obl}. The plot shows that the majority of the variation in $\Delta T$ is due to the mass-splitting encoded in $\Delta$. This is expected. The $T$ parameter is sensitive to isospin violation and thus constrains the splitting for $SU(2)_L$ multiplets. Viable benchmark points occur in the region with $\Delta \approx 0$, as seen in the plot, while larger mass-splittings can conflict with the constraints. 

The benchmark points include a range of values for the mass-splitting parameter $\Delta$, giving rise to the variation in Figure~\ref{fig:obl}. However,  in general, one can take the couplings $\lambda_{4,5}$ in the scalar potential sufficiently small to ensure the mass-splitting for $S^+$, $S^0$ and $A$ is consistent with oblique constraints. From the (technical) naturalness perspective,  arbitrarily small values of $\lambda_5$ are allowed, due to the enhanced lepton number symmetry for $\lambda_5\rightarrow0$.\footnote{In practice, the demand of viable neutrino masses gives a Yukawa coupling-dependent lower bound on $\lambda_5$.} Natural values of $\lambda_4$ are bounded from below by one-loop gauge contributions to the operator $|H^\dagger S|^2$. Consequently the mass splitting for components of $S$ is not expected to be smaller than the one-loop induced splitting, which is safely within the bounds. Thus, although  the oblique parameters can exclude some regions of parameter space, the constraints are readily evaded.

\begin{figure}[h]
\begin{center}
\includegraphics[width = 0.6\textwidth]{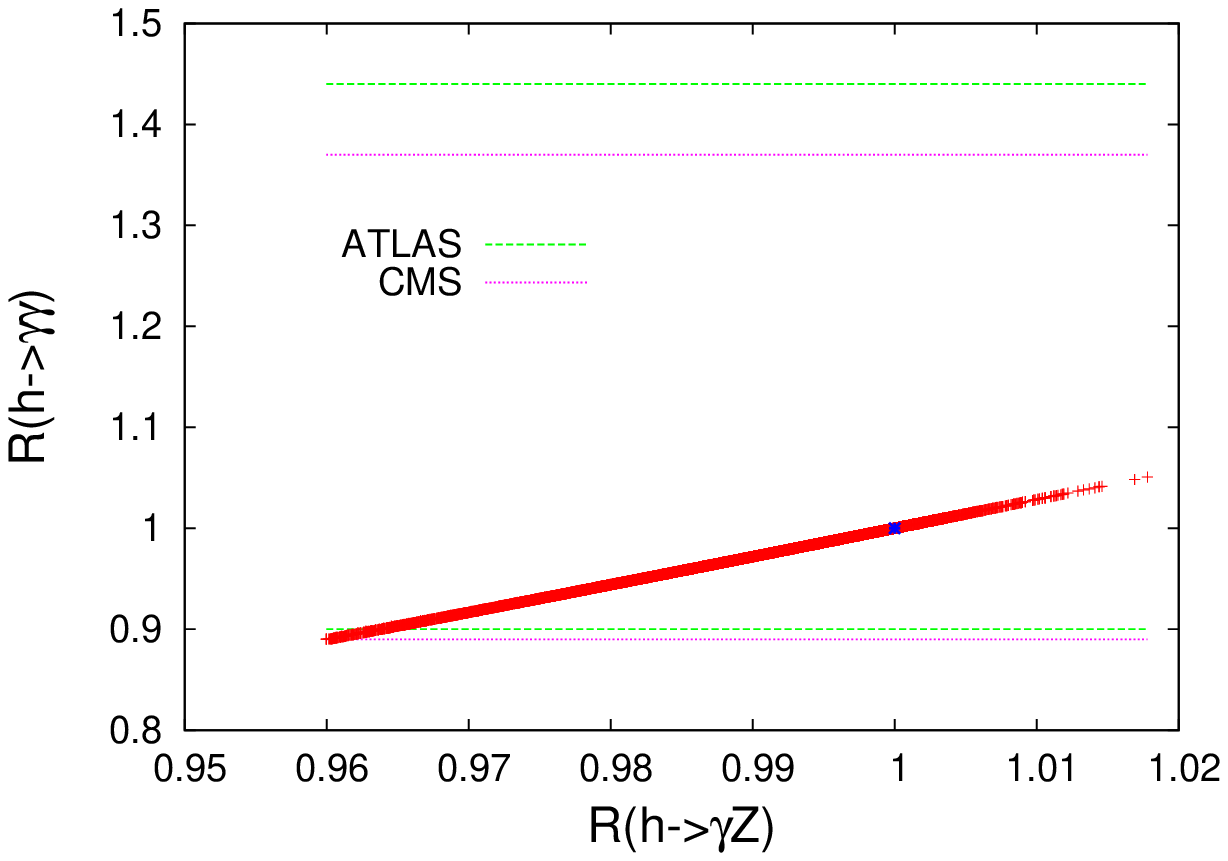}
\end{center}
\caption{Ratio of the widths for $h\rightarrow \gamma \gamma $ and $%
h\rightarrow \gamma Z$ relative to the SM values. The constraints
from ATLAS and CMS are shown. }
\label{fig:hYZ}
\end{figure}

The exotics in the model can also give  new
contributions to the Higgs decays $%
h\rightarrow \gamma Z$ and $h\rightarrow \gamma \gamma$. The ratio of the corresponding widths,
relative to the SM values, is plotted  in Figure~\ref{fig:hYZ}. One sees that the overwhelming majority of the benchmark points are consistent with 
constraints from ATLAS and CMS. Importantly, more-precise measurements by ATLAS and CMS during Run II of the LHC will provide further probes of the model. 

Before concluding, we note that our analysis reveals considerable differences between the SI scotogenic model and the standard (non-SI) scotogenic model. These relate primarily to the presence of the dilaton. The coupling between $\phi$ and the DM provides new annihilation channels for the sterile neutrino DM. This alleviates the need for larger Yukawa couplings $g_{i\alpha}$, normally required in the scotogenic model to generate the relic density, and reduces the tension with LFV constraints. However, the dilaton also permits new channels at direct-detection experiments making these constraints more severe for the SI  model. As a rough guide, one expects stronger LFV signals for the scotogenic model, and stronger direct-detection signals for the SI scotogenic model.


\section{Conclusion\label{sec:conc}}
In this work, we performed a detailed study of the minimal  SI  scotogenic model. Our analysis demonstrates the existence of   viable parameter space in which one obtains radiative electroweak symmetry breaking,  one-loop neutrino masses and a good DM candidate.  The model predicts a new scalar with $\mathcal{O}(\mathrm{GeV})$ mass. This field plays the dual roles of triggering electroweak symmetry breaking and sourcing lepton number symmetry violation. The model can give observable signals in LFV searches, direct-detection experiments, and precision searches for the Higgs decays $h\rightarrow \gamma\gamma$ and $h\rightarrow\gamma Z$. It also predicts a scalar doublet $S$, whose mass is expected to be $\lesssim $~TeV, within reach of collider experiments. The model is subject to strong constraints from direct-detection experiments;  viable parameter space was found for $M_\dm\lesssim10$~GeV and $M_\dm\gtrsim 200$~GeV, while intermediate values for $M_\dm$ appear excluded.
\section*{Acknowledgments\label{sec:ackn}}
AA is supported by the Algerian Ministry of Higher Education and
Scientific Research under the CNEPRU Project No D01720130042. KM is supported by the Australian Research Council.
\appendix


\end{document}